\documentclass[11pt,english,epsf]{scrartcl}
\usepackage{lmodern}
\usepackage[T1]{fontenc}
\usepackage[latin9]{inputenc}
\usepackage{geometry}
\usepackage{amsmath}
\usepackage{amssymb}

\newtheorem{theorem}{Theorem}

\newcommand {\dfn} {\stackrel{\Delta} {=}}
\newcommand {\exe} {\stackrel{\cdot} {=}}
\newcommand {\lexe} {\stackrel{\cdot} {\le}}
\newcommand {\gexe} {\stackrel{\cdot} {\ge}}

\newcommand {\bx} {\mbox{\boldmath $x$}}
\newcommand {\by} {\mbox{\boldmath $y$}}

\newcommand {\bE} {\mbox{\boldmath $E$}}

\newcommand {\bX} {\mbox{\boldmath $X$}}
\newcommand {\bY} {\mbox{\boldmath $Y$}}

\newcommand{\hP}{{\hat{P}}}
\newcommand{\tP}{{\tilde{P}}}
\newcommand{\hH}{{\hat{H}}}

\newcommand{\hI}{{\hat{I}}}
\newcommand{\tI}{{\tilde{I}}}
\newcommand{\calA}{{\cal A}}

\newcommand{\calC}{{\cal C}}

\newcommand{\calE}{{\cal E}}

\newcommand{\calI}{{\cal I}}

\newcommand{\calN}{{\cal N}}

\newcommand{\calT}{{\cal T}}

\newcommand{\calX}{{\cal X}}
\newcommand{\calY}{{\cal Y}}

\allowdisplaybreaks

\begin{document}
\thispagestyle{empty}
\title{List Decoding -- Random Coding Exponents and Expurgated Exponents\thanks{This 
research was supported by
the Israeli Science Foundation (ISF), grant no.\ 412/12.}}
\date{}
\author{Neri Merhav}
\maketitle

\begin{center}
Department of Electrical Engineering \\
Technion - Israel Institute of Technology \\
Technion City, Haifa 32000, ISRAEL \\
E--mail: {\tt merhav@ee.technion.ac.il}\\
\end{center}
\vspace{1.5\baselineskip}
\setlength{\baselineskip}{1.5\baselineskip}

\begin{center}
{\bf Abstract}
\end{center}
\setlength{\baselineskip}{0.5\baselineskip}
Some new results are derived concerning random coding error exponents
and expurgated exponents
for list decoding with a deterministic list size $L$. Two asymptotic regimes 
are considered, the {\it fixed list--size} regime, where $L$ is fixed independently of the block
length $n$, and the {\it exponential list--size}, where $L$ grows exponentially
with $n$. We first derive a general upper bound on the list--decoding average error
probability, which is suitable for both regimes. 
This bound leads to more specific
bounds in the two regimes. In the fixed list--size regime, the bound is
related to known bounds and we establish its exponential tightness.
In the exponential list--size regime, we establish
the achievability of the well known sphere packing lower bound.
Relations to guessing exponents are also provided. 
An immediate byproduct of our analysis in both regimes is
the universality of the maximum mutual information (MMI) list
decoder in the error exponent sense. 
Finally, we consider
expurgated bounds at low rates, both using Gallager's approach and the
Csisz\'ar--K\"orner--Marton approach, which is, in general better (at least
for $L=1$).
The latter expurgated bound, which
involves the notion of {\it multi--information}, is also modified to apply to
continuous alphabet channels, and in particular, to the Gaussian memoryless
channel, where the expression of the expurgated bound becomes quite explicit.

\vspace{0.2cm}


\setlength{\baselineskip}{2\baselineskip}
\newpage

\section{Introduction}

The concept of list decoding was first introduced independently by Elias
\cite{Elias57} and Wozencraft \cite{Wozencraft58} in the late fifties of the
previous century. A list decoder, rather than outputting a single estimate of
the transmitted message, produces a list of $L$ candidates, among which the final
`winner' will be eventually selected upon receiving additional information. This is
naturally applicable in 
concatenated coded communication systems, most notably, in wireless systems (see, e.g., \cite{SS89},
\cite{SS94}), where the list decoder corresponds to the inner code, and the
above--mentioned additional information is available to the outer decoder,
for example, information concerning the structure of the outer code or
contextual information that is present when not all the source redundancy has
been removed. Accordingly, the error event associated with a list decoder is
the event where the actual message that has been sent is not in the list.

While considerable work has been done on list decoding for codes with certain
algebraic/combinatorial structures 
(most notably, linear codes along with some subclasses 
of linear codes -- see, e.g., \cite{BMKF07}, \cite{Guruswami01} and many
references therein), with emphasis on algorithmic
issues, the main focus of this work is on Shannon--theoretic
issues, like achievable error exponents pertaining to list decoding.

A few words of background on list decoding error exponents are therefore in
order. First, a clear distinction should be made between two
different classes of list decoders, according to whether the list size is a
deterministic number or a random variable, which depends on the random 
channel output vector (see, e.g., \cite{Forney68}, \cite{Telatar97},
\cite{Telatar98} as well as many other later further developments).
In this paper, we confine ourselves to the former class, which in turn, is also
subdivided into two sub--classes with two
different asymptotic regimes: (i) the {\it fixed list--size regime}, where the 
deterministic list size $L$ is fixed, independently of the block length $n$, and (ii) the
{\it exponential list--size regime}, where $L$ is an exponential function
of the block length $n$, namely, $L=e^{\lambda n}$, for some fixed $\lambda > 0$, that is
independent of $n$. Both regimes will be considered in this paper.

First, regarding the fixed list--size regime, the extension of the ordinary random coding error exponent
analysis to account for list decoding with a fixed list size $L$, was
considered by both Gallager \cite[p.\ 538, Exercise 5.20]{Gallager68} and Viterbi and Omura
\cite[p.\ 215, Exercise 3.16]{VO79} simple and
straightforward enough to be left as an exercise for the reader.
In particular, for the ensemble of independent random selection of codewords
according to an i.i.d.\ distribution, in the above exercises it is shown that
the average probability of list--decoding error is upper bounded by an exponential
function, whose exponential decay rate is given by the same
expression as that of Gallager's random coding exponent, $E_{\mbox{\tiny r}}(R)$
\cite[eq.\
(5.6.16)]{Gallager68} ($R$ being the coding rate), except that the
interval of the maximization over the auxiliary parameter $\rho$ is expanded from
$[0,1]$ to
$[0,L]$ (thus $L=1$ recovers ordinary decoding as a
special case). Obviously, when $L$ exceeds the global maximizer of the same
expression over $[0,\infty)$, the resulting exponent coincides with the sphere--packing
exponent $E_{\mbox{\tiny sp}}(R)$ \cite[Theorem 5.8.1]{Gallager68}. Similarly
as in ordinary decoding, while the upper bound and the lower bound agree at high
rates, there is some gap at low rates and the random coding achievability
result can be improved at low rates by expurgation. Blinovsky
\cite{Blinovsky01} has shown that, in analogy to ordinary decoding, the
expurgated exponent of the fixed list--size regime at $R=0$ is tight in the sense that
there is also a lower bound of the same exponential rate.

The exponential list--size regime is interesting at least as much. In 1967, Shannon, Gallager and
Berlekamp \cite[Theorem 2]{SGB67} have established (among other results),
a lower bound 
on the probability of list decoding, which is meaningful also for the
exponential list--size regime. According to this lower bound, the probability of
list error cannot decay exponentially more rapidly than $e^{-nE_{\mbox{\tiny
sp}}(R-\lambda)}$, where again, $\lambda$ is the list--size exponent (see also
\cite[p.\ 196, Problem 27, part (b)]{CK81},
\cite[p.\ 179, Lemma 3.8.1]{VO79}). On the other hand, in \cite[p.\ 196,
Problem 27, part (a)]{CK81}, the reader is asked to show that an exponential
decay at the rate $e^{-nE_{\mbox{\tiny
r}}(R-\lambda)}$ is achievable (see also \cite[p.\ 3767, eq.\ (46)]{HSS10}).
One of our results is that a decay rate of $e^{-nE_{\mbox{\tiny
sp}}(R-\lambda)}$ is achievable, thus closing the gap between the upper bound
and the lower bound and fully characterizing the best achievable error exponent
function (reliability function for list decoding) in
the exponential list--size regime according to $E_{\mbox{\tiny
sp}}(R-\lambda)$.

In this paper, we contribute several additional results, both on the fixed
list--size
regime and the exponential list--size regime.
We first derive a general upper bound on the average list--decoding error
probability, pertaining to the ordinary ensemble of fixed composition codes,
i.e., the ensemble defined by an independent, random selection of each
codeword from a single type
class. This general bound is 
suitable for both regimes. 
The general upper bound leads to more
specific upper bounds in the two regimes. In the fixed list--size regime, our bound is
intimately related to the above mentioned bounds with the extended interval of
optimization  -- $[0,L]$, except that it corresponds to the ensemble of fixed
composition codes (rather than an i.i.d.\
codeword distribution). We also show that this random coding bound is
exponentially tight for the average code by deriving a compatible lower bound with the same
exponent, thereby extending the result in \cite{Gallager73} to list decoding.
In the exponential list size--regime, we establish, as mentioned
earlier, the achievability of $E_{\mbox{\tiny
sp}}(R-\lambda)$ and thereby close the gap between the lower bound and the
upper bound. An immediate byproduct of
these derivations is the universality of the maximum mutual information (MMI) list
decoder in the error exponent sense. 
The derivations involve
a random variable that is defined as the number of incorrect codewords whose
likelihood score exceed the score of the correct codeword. Accordingly, we
also provide bounds on the moments of this random variable, which are
intimately related to  
guessing exponents \cite{AM98a}, \cite{AM98b}. Finally, we consider
expurgated bounds at low rates, using both Gallager's approach
\cite[Section 5.7]{Gallager68}, \cite[Section 3.3]{VO79} and the
Csisz\'ar--K\"orner--Marton approach \cite[p.\ 185, problem 17]{CK81},
\cite{CKM77}. which is
in general better at least
for $L=1$ \cite{SPMMF13}.
The latter expurgated bound, which happens to
involve the notion of {\it multi--information},\footnote{Multi--information
\cite{SV98} is a 
natural generalization of the notion of 
mutual information that accommodates the statistical dependence
of more than two random
variables. It is defined as the sum of the marginal entropies of these random
variables minus their joint
entropy, or equivalently, as the Kullback--Leibler divergence between the joint distribution and
the product of marginals. To the best of our knowledge, in this paper, it is the first
occasion that the
multi--information plays a natural role in a concrete information--theoretic
formula.}
is also modified to apply to
continuous alphabet channels, and in particular, to the Gaussian memoryless
channel, where the expression of the expurgated bound becomes quite explicit.

The outline of the remaining part of this paper is as follows.
In Section 2, we establish notation conventions, formalize the
problem and provide some background and preliminaries. 
In Section 3, we first derive the general upper
bound on the average probability of list error for the
ordinary ensemble of fixed composition code (Subsection 3.1), and then particularize its analysis to
both the fixed list--size regime (Subsection 3.2) and
the exponential list--size regime (Subsection 3.3), and finally, we relate our
findings to the problem of guessing (Subsection 3.4). Section 4 is devoted to
expurgated exponents. Finally, in Section 5, we present a problem for future
work.

\section{Notation Conventions, Problem Formulation and Preliminaries}

\subsection{Notation Conventions}

Throughout the paper, random variables will be denoted by capital
letters, specific values they may take will be denoted by the
corresponding lower case letters, and their alphabets
will be denoted by calligraphic letters. Random
vectors and their realizations will be denoted,
respectively, by capital letters and the corresponding lower case letters,
both in the bold face font. Their alphabets will be superscripted by their
dimensions. For example, the random vector $\bX=(X_1,\ldots,X_n)$, ($n$ --
positive integer) may take a specific vector value $\bx=(x_1,\ldots,x_n)$
in $\calX^n$, the $n$--th order Cartesian power of $\calX$, which is
the alphabet of each component of this vector. 
Sources and channels will be denoted by the letter $P$ or $Q$, 
superscripted by the names of the relevant random variables/vectors and their
conditionings, if applicable, following the standard notation conventions,
e.g., $Q_X$, $P_{Y|X}$, and so on. When there is no room for ambiguity, these
subscripts will be omitted.
The probability of an event $\calE$ will be denoted by $\mbox{Pr}\{\calE\}$,
and the expectation
operator with respect to (w.r.t.) a probability distribution $P$ will be denoted by
$\bE_P\{\cdot\}$. Again, the subscript will be omitted if the underlying
probability distribution is clear from the context.
The entropy of a generic distribution $Q$ on $\calX$ will be denoted by
$H(Q)$. For two
positive sequences $a_n$ and $b_n$, the notation $a_n\exe b_n$ will
stand for equality in the exponential scale, that is,
$\lim_{n\to\infty}\frac{1}{n}\log \frac{a_n}{b_n}=0$. Similarly,
$a_n\lexe b_n$ means that
$\limsup_{n\to\infty}\frac{1}{n}\log \frac{a_n}{b_n}\le 0$, and so on.
The indicator function
of an event $\calE$ will be denoted by $\calI\{E\}$. The notation $[x]_+$
will stand for $\max\{0,x\}$.

The empirical distribution of a sequence $\bx\in\calX^n$, which will be
denoted by $\hat{P}_{\bx}$, is the vector of relative frequencies $\hat{P}_{\bx}(x)$
of each symbol $x\in\calX$ in $\bx$.
The type class of $\bx\in\calX^n$, denoted $\calT(\bx)$, is the set of all
vectors $\bx'$
with $\hat{P}_{\bx'}=\hat{P}_{\bx}$. When we wish to emphasize the
dependence of the type class on the empirical distribution $\hat{P}$, we
will denote it by
$\calT(\hat{P})$. Information measures associated with empirical distributions
will be denoted with `hats' and will be subscripted by the sequences from
which they are induced. For example, the entropy associated with
$\hat{P}_{\bx}$, which is the empirical entropy of $\bx$, will be denoted by
$\hat{H}_{\bx}(X)$. An alternative notation, following the conventions
described in the previous paragraph, is $H(\hP_{\bx})$.
Similar conventions will apply to the joint empirical
distribution, the joint type class, the conditional empirical distributions
and the conditional type classes associated with pairs (and multiples) of
sequences of length $n$. 
Accordingly, $\hP_{\bx\by}$ would be the joint empirical
distribution of $(\bx,\by)=\{(x_i,y_i)\}_{i=1}^n$,
$\calT(\bx,\by)$ or $\calT(\hP_{\bx\by})$ will denote
the joint type class of $(\bx,\by)$, $\calT(\bx|\by)$ will stand for
the conditional type class of $\bx$ given
$\by$, $\hH_{\bx\by}(X,Y)$ will designate the empirical joint entropy of $\bx$ and $\by$,
$\hH_{\bx\by}(X|Y)$ will be the empirical conditional entropy,
$\hI_{\bx\by}(X;Y)$ will
denote empirical mutual information, and so on.

\subsection{Problem Formulation}

Let $\{P(y|x)~x\in\calX,~y\in\calY\}$ be a matrix single--letter transition
probabilities of a discrete memoryless channel (DMC) 
with finite input and output alphabets, $\calX$ and
$\calY$, respectively. For a given $R> 0$, 
and a block length $n$, let $\calC=\{\bx_0,\ldots,\bx_{M-1}\}$, $M=e^{nR}+1$,
$\bx_m\in\calX^n$, $m=0,1,\ldots,M-1$, be a given codebook,
known to both the encoder and
decoder.\footnote{To avoid the need for rounding functions, we assume
throughout that $R$
is such that $e^{nR}$ is integer. Also, $M$ is defined as $e^{nR}+1$, rather
than $e^{nR}$, simply for reasons of convenience.}
We consider a list decoder that outputs a list of $L$ candidate
estimates of the transmitted message, 
based on the channel output $\by\in\calY^n$. We will be interested in two
asymptotic regimes. In the first regime, 
which will be referred to as the {\it fixed list--size regime},
$L$ is fixed and independent of $n$ and
in the second regime, 
which will be referred to as the {\it exponential list--size regime},
$L=e^{\lambda n}$, where $0< \lambda < R$ is a given
constant, independent of $n$. The figure of merit, in both cases, is the probability of list
error, namely, the probability that the correct message is not in the list.
Our focus will be on achievable error exponents associated with the
probability of list error as functions of $(R,L)$ in the fixed list--size
regime, or as functions of $(R,\lambda)$ in the exponential list--size regime.

As usual, the main mechanism for deriving achievability results will be random
coding. The random coding ensemble will be defined by independent random
selection of each codeword according to a probability distribution $P(\bx)$,
which, unless specified otherwise, will be the uniform distribution across
a given type class $\calT(Q)$ (except for a few places, where it will be the
product measure $\prod_{i=1}^nQ(x_i)$).
Once selected, the codebook is revealed to both
the encoder and the decoder.

The discussion in the subsections 2.3 and 3.1 is non--asymptotic -- it
applies to any two positive integers $n$ and $L\le e^{nR}$. 
This means, of course, that in these subsections, it is immaterial if we consider
the fixed list--size regime or the exponential list--size regime. 

\subsection{Preliminaries}

We begin from the fundamental fact that, as intuition suggests,
the optimal list decoder generates the list
according to the $L$ largest likelihood values $\{P(\by|\bx_m)\}$. This
follows from the following simple consideration, which is a fairly
simple extension of the one used to prove the optimality of the maximum likelihood (ML)
decoder in ordinary decoding ($L=1$). For a given unordered list of $L$
distinct integers $m_1,\ldots,m_L$, all in the range $\{0,1,\ldots,M-1\}$, let
$\Omega(m_1,\ldots,m_L)\subseteq\calY^n$
denote the region where the list decoder outputs the messages
$m_1,\ldots,m_L$. Obviously, the $\left(\begin{array}{cc}
M\\L\end{array}\right)$ different regions $\{\Omega(m_1,\ldots,m_L)\}$ form a
partition of $\calY^n$.
For a given $\by$, let $m_1^*(\by),\ldots m_L^*(\by)$
achieve the $L$ highest rankings of $P(\by|\bx_m)$.
Then, the probability of correct list decoding 
(i.e., the probability that the correct message is in the list)
is given by
\begin{eqnarray}
P_c&=&\frac{1}{M}\sum_{m=1}^M\sum_{m_1,\ldots,m_L}\left[\sum_{i=1}^L\delta(m_i-m)\right]
\sum_{\by\in\Omega(m_1,\ldots,m_L)}P(\by|\bx_m)\nonumber\\
&=&\frac{1}{M}\sum_{m_1,\ldots,m_L}\sum_{\by\in\Omega(m_1,\ldots,m_L)}
\sum_{i=1}^L\sum_{m=1}^M
\delta(m_i-m)P(\by|\bx_m)\nonumber\\
&=&\frac{1}{M}\sum_{m_1,\ldots,m_L}\sum_{\by\in\Omega(m_1,\ldots,m_L)}
\sum_{i=1}^LP(\by|\bx_{m_i})\nonumber\\
&\le&\frac{1}{M}\sum_{m_1,\ldots,m_L}\sum_{\by\in\Omega(m_1,\ldots,m_L)}
\sum_{i=1}^LP(\by|\bx_{m_i^*(\by)})\nonumber\\
&=&\frac{1}{M}\sum_{\by\in\calY^n}
\sum_{i=1}^LP(\by|\bx_{m_i^*(\by)})
\label{Pc}
\end{eqnarray}
and the inequality becomes an equality for the list decoder that outputs
$m_1^*(\by),\ldots,m_L^*(\by)$.

The following results, on random coding exponents for list decoding, are well
known. First, it is shown both in
Gallager \cite[p.\ 538, Exercise 5.20]{Gallager68} and Viterbi and Omura
\cite[p.\ 215, Exercise 3.16]{VO79}, that the average probability of list
error is upper bounded by
\begin{equation}
\label{vog}
\overline{P_e}\le \min_{0\le\rho\le
L}M^{\rho}
\sum_{\by\in\calY^n}\left[\sum_{\bx\in\calX^n}P(\bx)P(\by|\bx)^{1/(1+\rho)}\right]^{1+\rho},
\end{equation}
which is very similar to Gallager's well--known random coding bound for
ordinary decoding, except that here, the range of optimization of the
parameter $\rho$ is stretched from $[0,1]$ to $[0,L]$. When the random coding
distribution $P(\bx)$ is i.i.d.\ with a single--letter marginal $Q$, 
and the fixed list--size regime is considered,
this is, of course, an exponential function whose 
exponential rate is
\begin{equation}
\label{rceiid}
E_{\mbox{\tiny r}}(R,L)=\sup_{0\le\rho\le L}\sup_Q[E_0(\rho,Q)-\rho R],
\end{equation}
where $E_0(\rho)$ is the well--known Gallager function
\begin{equation}
E_0(\rho,Q)=-\ln\left(\sum_{y\in\calY}\left[\sum_{x\in\calX}
Q(x)P(y|x)^{1/(1+\rho)}\right]^{1+\rho}\right).
\end{equation}
Thus, $E_{\mbox{\tiny r}}(R,1)$ is the ordinary random coding error exponent,
which will also be denoted by $E_{\mbox{\tiny r}}(R)$, following traditional
notation conventions. 

Concerning the exponential list--size regime, where $L=e^{\lambda n}$, Shannon,
Gallager and Berlekamp \cite{SGB67} have proved that the exponential rate of
the list error probability cannot be faster than $E_{\mbox{\tiny
sp}}(R-\lambda)$, where $E_{\mbox{\tiny
sp}}(R)$ is the sphere--packing exponent, defined as
\begin{equation}
\label{gsp}
E_{\mbox{\tiny
sp}}(R)=\sup_{\rho\ge 0}\sup_Q[E_0(\rho,Q)-\rho R]
\end{equation}
or, equivalently,\footnote{In \cite{CK81}, see eq.\ (5.19) on page 166 and compare with Problem 23 on
page 192. See also the next footnote below.} as 
\begin{equation}
\label{csp}
E_{\mbox{\tiny
sp}}(R)=\sup_Q\inf_{\{\tilde{P}_{Y|X}:~\tilde{I}(X;Y)\le
R\}}D(\tilde{P}_{Y|X}\|P_{Y|X}|Q),
\end{equation}
where $\tilde{I}(X;Y)$ is the mutual information induced by $X\sim Q$ and
$\tilde{P}_{Y|X}$ and
\begin{equation}
D(\tilde{P}_{Y|X}\|P_{Y|X}|Q)\dfn\sum_{x\in\calX}Q(x)\sum_{y\in\calY}\tilde{P}_{Y|X}(y|x)\ln
\frac{\tilde{P}_{Y|X}(y|x)}{P_{Y|X}(y|x)}.
\end{equation}
In \cite[Problem 27]{CK81}, the reader is asked to prove that $E_{\mbox{\tiny
r}}(R-\lambda)$ is achievable.

\section{Upper Bounds for an Ordinary Ensemble of Fixed Composition Codes}

In this section, we first derive the general upper
bound on the average probability of list error for the
ordinary ensemble of fixed composition codes defined above (Subsection 3.1), and then
particularize its analysis to
both the fixed list--size regime (Subsection 3.2) and
the exponential list--size regime (Subsection 3.3). Finally, we relate our
findings to the problem of guessing (Subsection 3.4). 

\subsection{A General Upper Bound}

The following theorem holds for every positive integer $n$ and every $L\le
e^{nR}$.

\begin{theorem}
Consider the random coding ensemble of rate $R$ codes for a
DMC $\{P(y|x),~x\in\calX,~y\in\calY\}$, as described in Subsection 2.2.
The average probability of list error, 
$\overline{P_e}$, associated with the
optimal list decoder, is upper bounded by
\begin{equation}
\label{generalub}
\overline{P_e}\le
\sum_{\bx.\by}P(\bx)P(\by|\bx)\exp\left\{-nL\left[\hat{I}_{\bx\by}(X;Y)+\frac{\ln
L}{n}-R-\delta_n-\frac{1}{n}\right]_+\right\},
\end{equation}
where $P(\bx)$ is the uniform distribution
over $\calT(Q)$ and $\delta_n=O((\log n)/n)$.
\end{theorem}

Eq.\ (\ref{generalub}) serves as our general upper bound.
It can be
further analyzed, using the method of types, both
in the fixed list--size regime and in the
in the exponential list--size regime.
These analyses will be carried out in the following
two subsections, respectively. Note that the parameter $L$ appears twice in
the right--hand side (r.h.s.) of eq.\ (\ref{generalub}). In the fixed list--size regime, the
first occurrence of $L$ (outside the square brackets) 
will be the important one, whereas in the exponential
list--size regime, it will be the second occurrence of $L$ that will play the
more important role. The remaining part of this subsection is devoted to the
proof of Theorem 1.

\noindent
{\it Proof.}
For the purpose of deriving an upper bound, it is
legitimate to analyze a sub--optimal list decoder that outputs the $L$
messages with the $L$ highest values of the empirical mutual information
$\hI_{\bx_m\by}(X;Y)$, that is, the {\it maximum mutual information (MMI) list
decoder}. Without loss of generality, assume that $\bx_0$ was transmitted
and $\by$ was received. The average probability of list error, for a given
$(\bx_0,\by)$, denoted $\overline{P_e(\bx_0,\by)}$, is the probability of list
error which accounts for the randomness of all
other $M-1$ codewords. 
The overall average probability of list error, $\overline{P_e}$,
is the expectation of $\overline{P_e(\bX_0,\bY)}$ w.r.t.\ $P(\bx_0)P(\by|\bx_0)$,
where $P(\bx_0)$ is the uniform distribution across $\calT(Q)$.
For the given $\bx_0$ and $\by$, we then have
\begin{equation}
\overline{P_e(\bx_0,\by)}=\sum_{\ell=L}^M\left(\begin{array}{cc}
M\\ \ell\end{array}\right)[q(\bx_0,\by)]^\ell[1-q(\bx_0,\by)]^{M-\ell},
\end{equation}
where 
\begin{equation}
q(\bx_0,\by)=\frac{|\calT(Q)\cap\{\bx:~\hI_{\bx\by}(X;Y)\ge
\hI_{\bx_0\by}(X;Y)\}|}{|\calT(Q)|}.
\end{equation}
An alternative representation is the following: For a given $(\bx_0,\by)$,
let the random variable $N(\bx_0,\by)$ be defined as
\begin{equation}
N(\bx_0,\by)=\sum_{m=1}^{M-1}\calI\{\hI_{\bx_m\by}(X;Y)\ge
\hI_{\bx_0\by}(X;Y)\}.
\end{equation}
Then,
\begin{equation}
\overline{P_e(\bx_0,\by)}=\mbox{Pr}\{N(\bx_0,\by)\ge L\},
\end{equation}
and 
\begin{equation}
\overline{P_e}=\mbox{Pr}\{N(\bX_0,\bY)\ge L\},
\end{equation}
where it should be understood that the random variable $N(\bX_0,\bY)$
incorporates also the randomness of $\bX_0$ and $\bY$, in addition to the
randomness of $\{\bX_1,\ldots,\bX_{M-1}\}$, unlike $N(\bx_0,\by)$, for which
$\bx_0$ and $\by$ are fixed and only $\{\bX_1,\ldots,\bX_{M-1}\}$ are random.
By the method of types \cite{CK81},
\begin{eqnarray}
q(\bx_0,\by)&=&\sum_{\calT(\bx|\by)\subseteq \calT(Q):~\hI_{\bx\by}(X;Y)\ge
\hI_{\bx_0\by}(X;Y)}
\frac{|\calT(\bx|\by)|}{|\calT(\bx)|}\nonumber\\
&\le& \sum_{\calT(\bx|\by)\subseteq \calT(Q):~\hI_{\bx\by}(X;Y)\ge
\hI_{\bx_0\by}(X;Y)}
e^{-n[\hat{I}_{\bx\by}(X;Y)-\delta_n/2]}\nonumber\\
&\le& e^{-n[\hat{I}_{\bx_0\by}(X;Y)-\delta_n]}\dfn q'(\bx_0,\by),
\end{eqnarray}
where $\delta_n=O((\log n)/n)$ is a term that accounts for the normalized
logarithm of the number of different types classes of sequences of length $n$.
Obviously, $\mbox{Pr}\{N(\bx_0,\by)\ge L\}$ is a monotonically non--decreasing
function of $q(\bx_0,\by)$, and so 
\begin{equation}
\overline{P_e(\bx_0,\by)}\le
\sum_{\ell=L}^M\left(\begin{array}{cc}
M\\ \ell\end{array}\right)[q'(\bx_0,\by)]^\ell[1-q'(\bx_0,\by)]^{M-\ell}
\dfn \mbox{Pr}\{N'(\bx_0,\by)\ge L\}
\end{equation}
where $N'(\bx_0,\by)$ is defined as a Bernoulli random variable of $M-1$
independent trials and a probability of success $q'(\bx_0,\by)$.
Now, for a given $(\bx_0,\by)$, if $q'(\bx_0,\by) < L/(M-1)$,
the event $\{N'(\bx_0,\by)\ge L\}$ is a large deviations event, otherwise it occurs
with high probability.
Accordingly, the Chernoff bound on $\mbox{Pr}\{N(\bx_0,\by)\ge L\}$ is as
follows.
\begin{eqnarray}
\label{ub}
\overline{P_e(\bx_0,\by)}&\le&\left\{\begin{array}{ll}
\exp\{-MD(\frac{L}{M}\|q'(\bx_0,\by)\} & q'(\bx_0,\by)<L/(M-1)\\
1 & q'(\bx_0,\by)\ge L/(M-1)\end{array}\right.\nonumber\\
&=&\left\{\begin{array}{ll}
\exp\{-MD(\frac{L}{M}\|e^{-n[\hI_{\bx_0\by}(X;Y)-\delta_n]})\} &
\hI_{\bx_0,\by}(X;Y))>R-\frac{\ln L}{n}+\delta_n\\
1 & \hI_{\bx_0,\by}(X;Y))\le R-\frac{\ln L}{n}+\delta_n\end{array}\right.
\end{eqnarray}
where $D(a\|b)$, for $a,b\in[0,1]$, is the binary divergence function, that is
\begin{equation}
D(a\|b)=a\ln\frac{a}{b}+(1-a)\ln\frac{1-a}{1-b}.
\end{equation}
Now, for $a\ge b$, the following inequality is proved in \cite[pp.\ 167--168]{Merhav09}:
\begin{equation}
D(a\|b)\ge a\left[\ln\frac{a}{b}-1\right]_+.
\end{equation}
Thus, the first line of (\ref{ub}) is further upper bounded by
\begin{eqnarray}
& &\exp\left\{-e^{nR}Le^{-nR}\left[\ln\left(\frac{Le^{-nR}}
{\exp\{-n[\hat{I}_{\bx_0\by}(X;Y)-\delta_n]\}}\right)-1\right]_+\right\}\nonumber\\
&=&\exp\left\{-nL\left[\hat{I}_{\bx_0\by}(X;Y)+\frac{\ln
L}{n}-R-\delta_n-\frac{1}{n}\right]_+\right\},
\end{eqnarray}
and so, the upper bound on $\overline{P_e(\bx_0,\by)}$ can eventually
be presented, for every $(\bx_0,\by)$, as
\begin{equation}
\overline{P_e(\bx_0,\by)}\le
\exp\left\{-nL\left[\hat{I}_{\bx_0\by}(X;Y)+\frac{\ln
L}{n}-R-\delta_n-\frac{1}{n}\right]_+\right\}.
\end{equation}
The overall average list--error probability is obtained, of course, by averaging over
the randomness of $\bX_0$ and $\bY$. This completes the proof of Theorem 1.

\subsection{The Fixed List--Size Regime}

In the fixed list--size regime, the term $\frac{\ln L}{n}$, in the exponent of 
eq.\ (\ref{generalub}), vanishes as $n$ grows without bound. Since $P(\bx_0)$
is the uniform distribution within $\calT(Q)$, it is easy to see, by using the
method of types, that eq.\
(\ref{generalub}) leads to an exponential upper bound $\overline{P_e}\lexe
e^{-nE(R,L,Q)}$, where $E(R,L,Q)$ is given by
\begin{equation}
\label{erlq}
E(R,L,Q)=\min_{\tilde{P}_{Y|X}}\{D(\tilde{P}_{Y|X}\|P_{Y|X}|Q)+L\cdot[\tilde{I}(X;Y)-R]_+\},
\end{equation}
where $\tilde{I}(X;Y)$ and
$D(\tilde{P}_{Y|X}\|P_{Y|X}|Q)$ are defined as in Subsection 2.3.
Of course, the best bound is obtained by selecting $Q$ so as to maximize
$E(R,L,Q)$. The resulting exponent is then obviously $E(R,L)=\max_QE(R,L,Q)$.

We would like to compare eq.\ (\ref{erlq}) with eq.\ (\ref{vog}). As mentioned
in Subsection 2.3, when $P(\bx)=\prod_{i=1}^nQ(x_i)$, for the optimal $Q$, 
the resulting exponent is
$E_{\mbox{\tiny r}}(R,L)$, as defined in eq.\ (\ref{rceiid}).
Let us now evaluate the exponential rate of the r.h.s.\ of eq.\ (\ref{vog}) for
the case where $P(\bx)$ is uniform within $\calT(Q)$, as defined in Subsection
2.2. Using the
method of types, this gives the following:\footnote{This chain of exponential equalities
can serve as the basis for the proof that the two expressions of
$E_{\mbox{\tiny sp}}(R)$ (in Subsection 2.3) are equivalent. By taking
$L\to\infty$, the second to the last line becomes (\ref{csp}). On the other
hand, when $P(\bx)$ is the $n$--fold product of $Q$, the resulting exponent
would be (\ref{gsp}). For a given $Q$, (\ref{csp}) cannot be smaller than
(\ref{gsp}) since $\calI\{\bx\in\calT(Q)\}/|\calT(Q)|\lexe
\prod_{i=1}^nQ(x_i)$ for all $\bx$. On the other hand, the optimum input
distribution that minimizes the r.h.s.\ of (\ref{vog}) for a DMC is a product
distribution (see, e.g., \cite[Theorem 3]{Merhav00}), and
therefore, for the optimum $Q$, (\ref{gsp}) cannot be smaller than
(\ref{csp}). Consequently, for the optimum $Q$, the two expressions must be equal.}
\begin{eqnarray}
& &\min_{0\le\rho\le
L}M^{\rho}\sum_{\by}\left[\sum_{\bx\in\calT(Q)}\frac{1}{|\calT(Q)|}\cdot 
P(\by|\bx)^{1/(1+\rho)}\right]^{1+\rho}\\
&\exe&\min_{0\le\rho\le
L}M^{\rho}
\sum_{\by}\left[\sum_{\calT(\bx|\by)\subseteq\calT(Q)}|\calT(\bx|\by)|e^{-n\hat{H}(Q)}
\cdot\exp\left\{\frac{n}{1+\rho}\hat{\bE}_{\bx\by}\ln
P(Y|X)\right\}\right]^{1+\rho}\\
&\exe&\min_{0\le\rho\le
L}M^{\rho}
\sum_{\by}\left[\max_{\calT(\bx|\by)\subseteq\calT(Q)}\exp\left\{-n\hat{I}_{\bx\by}(X;Y)
+\frac{n}{1+\rho}\hat{\bE}_{\bx\by}\ln
P(Y|X)\right\}\right]^{1+\rho}\\
&\exe&\min_{0\le\rho\le
L}M^{\rho}
\sum_{\by}\exp\left\{-n\min_{\hP_{X|Y}:~\hP_X=Q}[(1+\rho)\hat{I}_{\bx\by}(X;Y)-
\hat{\bE}_{\bx\by}\ln
P(Y|X)]\right\}\\
&\exe&\exp\left\{-n\max_{0\le\rho\le
L}\min_{\tilde{P}_{XY}:~\tilde{P}_X=Q}[(1+\rho)\tilde{I}(X;Y)-\tilde{H}(Y)-
\tilde{\bE}\ln
P(Y|X)-\rho R]\right\}\\
&=&\exp\left\{-n\max_{0\le\rho\le
L}\min_{\tilde{P}_{Y|X}}[\rho\tilde{I}(X;Y)-\tilde{H}(Y|X)-\tilde{\bE}\ln
P(Y|X)-\rho R]\right\}\\
&=&\exp\left\{-n\max_{0\le\rho\le
L}\min_{\tilde{P}_{Y|X}}(D(\tilde{P}_{Y|X}\|P_{Y|X}|Q)+\rho[\tilde{I}(X;Y)-R])\right\}\\
&=&\exp\left\{-n\min_{\tilde{P}_{Y|X}}\max_{0\le\rho\le
L}(D(\tilde{P}_{Y|X}\|P_{Y|X}|Q)+\rho\cdot[\tilde{I}(X;Y)-R])\right\}\\
&=&\exp\left\{-n\min_{\tilde{P}_{Y|X}}
(D(\tilde{P}_{Y|X}\|P_{Y|X}|Q)+L\cdot[\tilde{I}(X;Y)-R]_+)\right\}\\
&=&e^{-nE(R,L,Q)},
\end{eqnarray}
where $\hat{\bE}_{\bx\by}\{\cdot\}$ and 
$\tilde{\bE}\{\cdot\}$ 
denote the expectation operators w.r.t.\ 
$\hP_{\bx\by}$ and $\tilde{P}_{XY}$, respectively, and $\tilde{H}(Y)$ and $\tilde{H}(Y|X)$ denote the
unconditional and the conditional output entropy induced by $\tilde{P}_{XY}$. Here,
we have used the fact that the minimization over $\tilde{P}_{Y|X}$ and the
maximization over $\rho$ are commutative since the objective function is
convex--concave. We see that
the two upper bounds
are exponentially equivalent in the fixed list--size regime,
for the ensemble of independent random selection within a given type class.
For $L\to\infty$, this is equivalent
$\min\{D(\tilde{P}_{Y|X}\|P_{Y|X}|Q):~\tilde{I}(X;Y)\le R\}$, which is the
sphere--packing exponent $E_{sp}(R)$. In fact, it is achieved for all $L\ge L_c$,
where $L_c$ is the smallest value of $L$ for which the minimizing
$\tilde{P}_{Y|X}$ of
$D(\tilde{P}_{Y|X}\|P_{Y|X}|Q)+L\cdot[\tilde{I}(X;Y)-R]_+$
achieves $\tilde{I}(X;Y)\le R$ (think of $L$ as a Lagrange multiplier).

We next show that this bound is also
exponentially tight for the average code (thus
extending the main result of \cite{Gallager73} from ordinary decoding to list
decoding), by deriving a compatible lower bound on the average error
probability. Consider now the optimum list decoder, that outputs the $L$ most
likely messages. First, the probability that a single incorrect codeword will receive
a likelihood score higher than that of the correct message, for a given
$\bx_0$ and $\by$, is lower bounded as follows:
\begin{eqnarray}
q_0(\bx_0,\by)&\dfn&\sum_{\calT(\bx|\by)\subset\calT(Q):~P(\by|\bx)\ge P(\by|\bx_0)}
\frac{|T(\bx|\by)|}{|\calT(Q)|}\\
&\ge&\frac{|T(\bx_0|\by)|}{|\calT(Q)|}\\
&\ge&\exp\{-n[\hI_{\bx_0\by}(X;Y)+\delta_n]\}\dfn q_0'(\bx_0,\by).
\end{eqnarray}
Let $N''(\bx_0,\by)$ be a Bernoulli random variable with $M-1$ independent trials and
probability of success $q_0'(\bx_0,\by)$. Then,
\begin{eqnarray}
\overline{P_e}&\ge&
\mbox{Pr}\{N''(\bX_0,\bY)\ge L\}\\
&=&\mbox{Pr}\{N''(\bX_0,\bY)\ge
L,~\hI_{\bx_0\by}(X;Y)>R-\delta_n\}+\nonumber\\
& &\mbox{Pr}\{N''(\bX_0,\bY)\ge L|\hI_{\bx_0\by}(X;Y)\le
R-\delta_n\}\cdot\mbox{Pr}\{\hI_{\bx_0\by}(X;Y)\le
R-\delta_n\}\\
&\exe&\mbox{Pr}\{N''(\bX_0,\bY)\ge L,~\hI_{\bx_0\by}(X;Y)>R-\delta_n\}+
\mbox{Pr}\{\hI_{\bx_0\by}(X;Y)\le R-\delta_n\}\\
&\exe&\mbox{Pr}\{N''(\bX_0,\bY)\ge L,~\hI_{\bx_0\by}(X;Y)>R-\delta_n\}+\nonumber\\
& &\exp\left\{-n\min_{\tilde{P}_{Y|X}:~\tilde{I}(X;Y)\le R}D(\tilde{P}_{Y|X}\|P_{Y|X}|Q)\right\},
\label{lowerbound}
\end{eqnarray}
where we have used fact that
given $\hI_{\bx_0\by}(X;Y)\le R-\delta_n$, the event
$\{N''(\bX_0,\bY)\ge L\}$ occurs with high probability.
As for the first term on the right--most side of (\ref{lowerbound}), consider first a given
$(\bx_0,\by)$ with $\hI_{\bx_0\by}(X;Y)> R-\delta_n$ and let
$\hat{N}(\bx_0,\by)$ be a Bernoulli random variable with $e^{n(R-\delta_n)}$ independent
trials and
probability of success $q_0'(\bx_0,\by)$. Then,
\begin{eqnarray}
\mbox{Pr}\{N''(\bx_0,\by)\ge L\}&\ge&
\mbox{Pr}\{\hat{N}(\bx_0,\by)\ge L\}\\
&\ge&\mbox{Pr}\{\hat{N}(\bx_0,\by)=L\}\\
&=&\left(\begin{array}{cc} e^{n(R-\delta_n)} \\ L\end{array}\right)\cdot
q_0'(\bx_0,\by)^L[1-q_0'(\bx_0,\by)]^{e^{n(R-\delta_n)}-L}\\
&\ge&\left(\begin{array}{cc} e^{n(R-\delta_n)} \\ L\end{array}\right)\cdot
q_0'(\bx_0,\by)^L[1-e^{-n[R-\delta_n]}]^{e^{n(R-\delta_n)}}\\
&\exe&e^{nRL}\cdot
\exp\{-nL\hI_{\bx_0\by}(X;Y)\}\\
&=&\exp\{-nL[\hI_{\bx_0\by}(X;Y)-R]\}.
\end{eqnarray}
Thus,
\begin{eqnarray}
& &\mbox{Pr}\{N''(\bX_0,\bY)\ge L,~\hI_{\bX_0\bY}(X;Y)>R-\delta_n\}\\
&\gexe&
\sum_{(\bx_0,\by):~\hI_{\bx_0\by}(X;Y)>R-\delta_n}P(\bx_0,\by)\cdot
\exp\{-nL[\hI_{\bx_0\by}(X;Y)-R]\}\\
&\exe&\exp\left\{-n\min_{\tilde{P}_{Y|X}:~\tilde{I}(X;Y)>R}
(D(\tilde{P}_{Y|X}\|P_{Y|X}\|Q)+L[\tilde{I}(X;Y)-R])\right\}.
\end{eqnarray}
Combining this with the second term on the right--most side of
(\ref{lowerbound}),
the overall exponent becomes
the minimum between
$$\min_{\tilde{P}_{Y|X}:~\tilde{I}(X;Y)\le
R}D(\tilde{P}_{Y|X}\|P_{Y|X}|Q)$$
and
$$\min_{\tilde{P}_{Y|X}:~\tilde{I}(X;Y)>R}(D(\tilde{P}_{Y|X}\|P_{Y|X}\|Q)+L[\tilde{I}(X;Y)-R]),$$
namely,
$$\min_{\tilde{P}_{Y|X}}\{D(\tilde{P}_{Y|X}\|P_{Y|X}\|Q)+L\cdot [\tilde{I}(X;Y)-R]_+\},$$
which is again exactly $E(R,L,Q)$ of eq.\ (\ref{erlq}). Thus, the upper bound (\ref{vog}) and
our fixed list--size bound (\ref{erlq}) are
equivalent under the random coding ensemble of fixed composition codes,
and they both give the exact random coding exponent for the average code.
Also, since the upper bound was obtained by the MMI list decoder and the compatible
lower bound applies to the optimal, ML list decoder, we have also proved, as a
byproduct, the universal
optimality of the MMI list decoder in the error exponent sense.

\subsection{The Exponential List--Size Regime}

In the exponential list--size regime, $L=e^{\lambda n}$. 
Now, let $\epsilon > 0$ be arbitrarily small, and 
define $\calE$ to be the set of all $\{(\bx,\by)\}$ for 
which 
\begin{equation}
\hat{I}_{\bx\by}(X;Y)+\lambda-R
-\delta_n-\frac{1}{n}\ge\epsilon.
\end{equation}
Then, eq.\ (\ref{generalub}) is averaged as
follows:
\begin{eqnarray}
\overline{P_e}&\lexe&\sum_{\bx,\by}P(\bx)P(\by|\bx)
\exp\left\{-ne^{n\lambda}\left[\hat{I}_{\bx\by}(X;Y)+\lambda
-R-\delta_n-\frac{1}{n}\right]_+\right\}\\
&=&\sum_{(\bx,\by)\in\calE}P(\bx)P(\by|\bx)
\exp\left\{-ne^{n\lambda}\left[\hat{I}_{\bx\by}(X;Y)+\lambda
-R-\delta_n-\frac{1}{n}\right]_+\right\}+\nonumber\\
& &+\sum_{(\bx_0,\by)\in\calE^c}P(\bx)P(\by|\bx)
\exp\left\{-ne^{n\lambda}\left[\hat{I}_{\bx\by}(X;Y)+\lambda
-R-\delta_n-\frac{1}{n}\right]_+\right\}\\
&\le& \exp\{-\epsilon e^{\lambda
n}\}+\sum_{(\bx,\by)\in\calE^c}P(\bx)P(\by|\bx).
\end{eqnarray}
Now, the first term  decays
double--exponentially and hence is negligible. Therefore,
$\overline{P_e}$ is dominated by the second term. 
Using the method of types, the arbitrariness of $\epsilon$, and the
fact that $\delta_n$ and $1/n$ vanish as $n\to\infty$,
the exponent of the second term is easily found to be
$$\min_{\{\tP_{Y|X}:~\tI(X;Y)\le R-\lambda\}}D(\tP_{Y|X}\|P_{Y|X}|Q),$$
whose maximum over $Q$ 
is $E_{\mbox{\tiny sp}}(R-\lambda)$, as mentioned in Subsection 2.3.
Thus, we have shown that the Shannon--Gallager--Berlekamp bound \cite[Theorem
2]{SGB67} is actually achievable across the whole relevant range of rates, 
$(\lambda,C+\lambda)$, where $C$ is the channel capacity.
Once again, we see that in the exponential list--size regime too,
the MMI list decoder is universal in the error exponent in sense.


\subsection{Relation to the Guessing Problem} 

It is interesting to look at moments of the random variable $N(\bX_0,\bY)$
(or any of the other Bernoulli random variables that we have define earlier).
Consider the following lower bound. Let $\epsilon > 0$ be arbitrarily small.
Then,
\begin{eqnarray}
\bE\{N(\bX_0,\bY)^\rho\}&=&\sum_{L=1}^ML^\rho\cdot\mbox{Pr}\{N(\bX_0,\bY)=L\}\\
&\ge&\sum_{i=0}^{R/\epsilon-1}e^{in\epsilon\rho}\cdot\mbox{Pr}\{e^{ni\epsilon}\le
N(\bX_0,\bY)<e^{n(i+1)\epsilon}\}\\
&\gexe&\sum_{i=0}^{R/\epsilon-1}e^{in\epsilon\rho}\cdot e^{-nE_{\mbox{\tiny
sp}}(R-i\epsilon)}\\
&\exe&\exp\left\{n\max_i[i\epsilon\rho-E_{\mbox{\tiny
sp}}(R-i\epsilon)]\right\}.
\end{eqnarray}
Since $\epsilon> 0$ is arbitrarily small, one can take the limit $\epsilon\to 0$,
and obtain
\begin{eqnarray}
\liminf_{n\to\infty}\frac{\ln\bE\{N(\bX_0,\bY)^\rho\}}{n}&\ge&
\sup_{0<\lambda< R}[\rho\lambda-E_{\mbox{\tiny
sp}}(R-\lambda)].
\end{eqnarray}
Let $\varrho(R)$ denote the achiever of 
$E_{\mbox{\tiny
sp}}(R)$ in eq.\ (\ref{gsp}). 
Then obviously, $\dot{E}_{\mbox{\tiny sp}}(R)=-\varrho(R)$,
where $\dot{E}_{\mbox{\tiny sp}}(R)$ denotes the derivative of
$E_{\mbox{\tiny sp}}(R)$. Thus,
the supremum is achieved by $\lambda^*=R-\varrho^{-1}(\rho)$,
provided that $\varrho^{-1}(\rho)\le R$, namely, $\rho\ge \varrho(R)$.
In this case, the exponential lower bound becomes
\begin{eqnarray}
\rho[R-\varrho^{-1}(\rho)]-E_{\mbox{\tiny sp}}(\varrho^{-1}(\rho))
&=&\rho R-\rho \varrho^{-1}(\rho)-E_0(\rho)+\rho \varrho^{-1}(\rho)\\
&=&\rho R-E_0(\rho).
\end{eqnarray}
When $\rho\ge \varrho(R)$, the supremum is achieved at $\lambda=0$
and the result is $-E_{\mbox{\tiny sp}}(R)$.
In summary, then
\begin{equation}
\liminf_{n\to\infty}\frac{\ln\bE\{N(\bX,\bY)^\rho\}}{n}\ge
\left\{\begin{array}{ll}
-E_{\mbox{\tiny sp}}(R) & \rho \le \varrho(R)\\
\rho R-E_0(\rho) & \rho > \varrho(R)\end{array}\right.
\label{nxy}
\end{equation}
This result is intimately related to those on guessing exponents \cite{AM98a},
\cite{AM98b}. 
In the guessing problem, the decoder of a coded communication system submits,
upon receiving the channel output $\by$, a sequence of guesses,
$\bx_{m_1},\bx_{m_2},\ldots$ concerning the
transmitted message until it hits the correct message. Let $G(\bx_0|\by)$
denote the number of guesses when the transmitted codeword is $\bx_0$ and the
received vector is $\by$. The aim is to minimize
$\bE\{[G(\bX_0|\bY)]^\rho\}$ for a given $\rho> 0$, and the best guessing
strategy is, of course, according to decreasing likelihood scores, hence the
close relation to list decoding. In fact, by definition, the relation between
$N(\bx_0,\by)$ and $G(\bx_0|\by)$ is extremely simple:
$G(\bx_0|\by)=N(\bx_0,\by)+1$. While this seems like a very
minor difference, it should be noted $\bE\{G(\bX_0|\bY)^\rho\}$ can only have a
non--negative exponent since $G(\bx_0|\by)\ge 1$, whereas
$\bE\{N(\bX_0,\bY)^\rho\}$ may also have a negative exponent since there is a
high probability that $N(\bX_0,\bY)=0$ (the event of correct decoding in the
ordinary sense).
Since $G(\bX_0|\bY)\ge\max\{N(\bX_0,\bY),1\}$, the exponent of
$\bE\{G(\bX_0|\bY)^\rho\}$ is lower bounded by an exponential function whose
exponent is given by the positive part of the exponent of 
$\bE\{N(\bX_0,\bY)^\rho\}$, and so, some information is lost when one considers
the guessing exponent rather than the exponent of 
$\bE\{N(\bX_0,\bY)^\rho\}$. In other words, the latter is a more informative
measure.
Indeed, the second line of eq.\ (\ref{nxy}) agrees with the results on
guessing exponents in \cite{AM98b}.\footnote{The
results in \cite{AM98b} are for lossy joint source--channel coding, however, it
is easy to particularize them to pure channel coding by considering 
the binary symmetric source and allowing zero distortion.}
Our results thus far seem to indicate that
the following performance is achievable (provided that the input distribution
that attains $E_{\mbox{\tiny sp}}(R)$ is independent of $R$):
\begin{equation}
\liminf_{n\to\infty}\frac{\ln\bE\{N(\bX_0,\bY)^\rho\}}{n}\le
\left\{\begin{array}{ll}
-E_{\mbox{\tiny r}}(R) & \rho \le \varrho'(R)\\
\rho R-E_0(\rho) & \rho > \varrho'(R)\end{array}\right.
\end{equation}
where $\varrho'(R)$ is the solution to the equation
$E_{\mbox{\tiny r}}(R)=E_0(\rho)-\rho R$. Thus, at least for large values of
$\rho$, where the dominant value of $\lambda$ is large enough, the bound is
tight (just like in the guessing problem).

\section{Upper Bounds for an Expurgated Ensemble}

While in the exponential list--size regime, we have an exact characterization
of the list--decoding reliability function, as $E_{\mbox{\tiny sp}}(R-\lambda)$,
in the fixed list--size regime, there is some gap 
between the upper bound and the lower bound, 
$E_{\mbox{\tiny sp}}(R)$, at low rates, just like in
ordinary decoding. Although this problematic range of low rates shrinks as $L$ increases,
it is nevertheless still existent for every finite $L$. Like in ordinary decoding, one
the way to reduce this gap is by improving the upper bound at low rates using
expurgation. In this section, we discuss expurgated bounds for list decoding.

For a given code $\calC$, consider first the following union bound
on $P_{e|m}(\calC)$, the probability of list error,
given that message no.\ $m$ was transmitted.
\begin{eqnarray}
P_{e|m}(\calC)&=&\sum_{m_1,\ldots,m_L\ne m}
\sum_{\by\in\calY^n}P(\by|\bx_m)\prod_{i=1}^L\calI\{P(\by|\bx_{m_i})\ge
P(\by|\bx_m)\}\nonumber\\
&\le&\sum_{m_1,\ldots,m_L\ne m}\sum_{\by\in\calY^n}
P(\by|\bx_m)\prod_{i=1}^L\left[\frac{P(\by|\bx_{m_i})}{
P(\by|\bx_m)}\right]^{1/(L+1)}\nonumber\\
&=&\sum_{m_1,\ldots,m_L\ne m}\sum_{\by\in\calY^n}
\left[P(\by|\bx_m)\prod_{i=1}^LP(\by|\bx_{m_i})\right]^{1/(L+1)},
\end{eqnarray}
where summation over $\{m_1,\ldots,m_L\}$ is over all $L$--tuples of distinct
integers in $\{0,1,\ldots,M-1\}\setminus\{m\}$.

The first approach to the derivation of expurgated bounds is a straightforward
generalization of Gallager's approach \cite[Section 5.7]{Gallager68} (see also
\cite[Section 3.3]{VO79}). 
Using the same argumentation as in the derivation of the classical expurgated
bound in these references, it is apparent that there exists a code for which
the {\it maximal} probability of error satisfies, for every $s\ge 0$:
\begin{equation}
\max_mP_{e|m}(\calC)\le 
\left[\bE\left(\sum_{m_1,\ldots,m_L\ne m}\sum_{\by\in\calY^n}
\left[P(\by|\bX_m)\prod_{i=1}^LP(\by|\bX_{m_i})\right]^{1/(L+1)}\right)^s\right]^{1/s},
\end{equation}
where now the summation over $\{m_1,\ldots,m_L\}$ is across all $L$--tuples of
distinct integers in $\{0,1,\ldots,2M-1\}\setminus\{m\}$.
Now, assuming $s\in[0,1]$, the Jensen inequality can be used to obtain
\begin{eqnarray}
\max_mP_{e|m}(\calC)&\le&
\left[\bE\left(\sum_{m_1,\ldots,m_L\ne m}\sum_{\by\in\calY^n}
\left[P(\by|\bX_m)\prod_{i=1}^LP(\by|\bX_{m_i})\right]^{1/(L+1)}\right)^s\right]^{1/s}\\
&\le&\left[\bE\sum_{m_1,\ldots,m_L\ne m}\left(\sum_{\by\in\calY^n}
\left[P(\by|\bX_m)\prod_{i=1}^LP(\by|\bX_{m_i})\right]^{1/(L+1)}\right)^s\right]^{1/s}\\
&\le&\left[(2M)^L\sum_{\bx_0,\bx_1,\ldots,\bx_L\in\calX^n}
\prod_{i=0}^LQ(\bx_i)\left(\sum_{\by\in\calY^n}
\prod_{0=1}^LP(\by|\bx_i)^{1/(L+1)}\right)^s\right]^{1/s},
\end{eqnarray}
which can be further developed using the method of types,
for the ensemble of fixed composition codes,
or using Gallager's
technique, for the ensemble where the distribution of each codeword is the
$n$--fold product measure $Q$. In the latter case, the resulting
single--letter formula of the expurgated exponent is
\begin{equation}
E_{\mbox{\tiny ex}}^{\mbox{\tiny G}}(R,L)
=\sup_{\rho\ge
1}\sup_Q\left\{-\rho\ln\left[\sum_{x_0,x_1,\ldots,x_L\in\calX}\prod_{i=0}^LQ(x_i)
\left(\sum_{y\in\calY}\prod_{i=0}^L[P(y|x_i)^{1/(L+1)}\right)^{1/\rho}\right]-\rho
LR\right\},
\end{equation}
where $\rho=1/s$ and where the superscript ``G'' stands for ``Gallager''. At zero rate,
\begin{equation}
E_{\mbox{\tiny ex}}^{\mbox{\tiny
G}}(0,L)=-\sum_{x_0,x_1,\ldots,x_L}\prod_{i=0}^LQ(x_i)
\ln \left(\sum_y\prod_{i=0}^LP(y|x_i)^{1/(L+1)}
\right)
\end{equation}
has been shown by Blinovsky \cite{Blinovsky01} to be tight in the
sense that there is also a compatible lower bound with the same exponential rate.

Another approach is in the spirit of the methodology of Csisz\'ar, K\"orner and Marton
\cite[page 185, Problem 17]{CK81}, \cite{CKM77}, which for $L=1$ (ordinary
decoding) gives an exponent at least as large as that of Gallager, with
equality for the optimal choice of $Q$ \cite[p.\ 193, Problem 23(ii)]{CK81}, \cite{SPMMF13}.
To the best of our knowledge, the extension of the Csisz\'ar--K\"orner--Marton
(CKM) expurgated exponent to list decoding is new. 

Define the following function, which is an extension of the Bhattacharyya
distance from two to $L+1$ variables:
\begin{equation}
d(x_0,x_1,\ldots,x_L)=-\ln\left[\sum_{y\in\calY}\prod_{i=0}^LP(y|x_i)^{1/(L+1)}\right].
\end{equation}
For a random vector
$(X_0,X_1,\ldots,X_L)$ with a given joint distribution $P_{X_0X_1\ldots X_L}$,
let us define also the {\it multi--information} as
\begin{equation}
I(X_0;X_1;\ldots;X_L)=\sum_{i=0}^LH(X_i)-H(X_0,X_1,\ldots,X_L)=D(P_{X_0X_1\ldots
X_L}\|P_{X_0}\times P_{X_1}\times\ldots P_{X_L}),
\end{equation}
which is a natural extension of the mutual information as a measure of
joint dependence among several random variables \cite{SV98}. 
Our main result in this section is the following.theorem, the proof of which
appears in the Appendix.

\begin{theorem}
There exists a sequence of
rate--$R$ codes for which
\begin{equation}
\lim_{n\to\infty}\left[-\frac{\ln \max_m P_{e|m}(\calC)}{n}\right]\ge
E_{\mbox{\tiny ex}}^{\mbox{\tiny CKM}}(R,L),
\end{equation}
where
\begin{equation}
E_{\mbox{\tiny ex}}^{\mbox{\tiny CKM}}(R,L)\dfn \sup_Q\inf_{\{P_{X_0X_1\ldots
X_L}\in\calA(R,Q)\}}
[\bE
d(X_0,X_1,\ldots,X_L)+I(X_0;X_1;\ldots;X_L)]-LR,
\end{equation}
and
\begin{equation}
\calA(R,Q)\dfn\{P_{X_0X_1\ldots X_L}:~
I(X_0;X_1;\ldots;X_L)\le LR,~P_{X_0}=P_{X_1}=\ldots=P_{X_L}=Q\}.
\end{equation}
\end{theorem}
It is easy to see that, similarly as in the case $L=1$, here too,
$E_{\mbox{\tiny ex}}^{\mbox{\tiny CKM}}(R,L)$ 
is a monotonically decreasing, convex
function for
some range of small rates, and then 
at a certain critical rate, it becomes an affine function (with slope
$-L$) in the range of large $R$. In particular, the convex function in the
low--rate range is the ``distortion--rate function'' 
\begin{equation}
D(R)=\min_{P_{X_0X_1\ldots X_L}\in\calA(R,Q)}
\bE\{d(X_0,X_1,\ldots,X_L)\},
\end{equation}
and the critical rate is $R_{\mbox{\tiny crit}}=I^*(X_0;X_1;\ldots;X_L)/L$, where
$I^*(X_0;X_1;\ldots;X_L)$ is the multi--information pertaining to the
minimizer $P_{X_0X_1\ldots X_L}^*$ of $\bE
d(X_0,X_1,\ldots,X_L)+I(X_0;X_1;\ldots;X_L)$ over $\calA(\infty,Q)$. The affine part is given by the
straight line that is tangential to the curve $D(R)$ at $R=R_{\mbox{\tiny
crit}}$.
The special case $L=1$ obviously recovers the CKM
expurgated exponent for ordinary decoding.

A few comments concerning the continuous alphabet case are in order.
The proof technique (see Appendix) can be extended, in principle, to the continuous alphabet case
(with input constraints), provided that we can assess exponential growth rates
of volumes\footnote{See, e.g., \cite[Proof of Theorem 5]{AM98a}, \cite{Merhav93}, where
an extension of the method of types to continuous alphabet situations was used
extensively.} of groups of
$(L+1)$ $n$--vectors with every given value of
$d(\bx_0,\bx_1,\ldots,\bx_L)=\sum_{i=1}^nd(x_{0,i},x_{1,i},\ldots,x_{L,i})$
(within an arbitrarily tolerance $\epsilon$).
For concreteness,
consider the memoryless Gaussian channel, $Y_i=X_i+Z_i$,
where $Z_i\sim\calN(0,\sigma^2)$ are i.i.d., independent of
$\bX=(X_1,\ldots,X_n)$, and the random codewords $\{\bX_m\}$
are independently selected under the uniform distribution over the surface of
the $n$--dimensional hypersphere of radius $\sqrt{nS}$ ($S>0$ being the power).
Then,
\begin{eqnarray}
d(x_0,x_1,\ldots,x_L)
&=&-\ln\left[\int_{-\infty}^{+\infty}\mbox{d}y\cdot
(2\pi\sigma^2)^{-1/2}\exp\left\{-\frac{1}{2\sigma^2(L+1)}\sum_{i=0}^L(y-x_i)^2\right\}\right]\\
&=&\frac{L}{2\sigma^2(L+1)^2}\left[\sum_{i=0}^Lx_i^2-\frac{2}{L}\sum_{i<j}x_ix_j\right]
\end{eqnarray}
and for vectors of length $n$, we have
\begin{eqnarray}
d(\bx_0,\bx_1,\ldots,\bx_L)&=&\frac{L}{2\sigma^2(L+1)^2}\sum_{t=1}^n
\left[\sum_{i=0}^Lx_{i,t}^2-\frac{2}{L}\sum_{i<j}x_{i,t}x_{j,t}\right]\\
&=&\frac{L}{2\sigma^2(L+1)^2}
\left[n(L+1)S-\frac{2}{L}\sum_{i<j}\sum_{t=1}^nx_{i,t}x_{j,t}\right]\\
&=&\frac{nLS}{2\sigma^2(L+1)^2}
\left(L+1-\frac{2}{L}\sum_{i<j}\rho_{ij}\right),
\end{eqnarray}
where $\rho_{ij}=\sum_{t=1}^nx_{t,i}x_{t,j}/(nS)$.
Thus, $d(\bx_0,\bx_1,\ldots,\bx_L)$ is completely determined by the empirical correlations
$\{\rho_{ij}\}$. Therefore,
in order to modify the proof of Theorem 2 to the continuous case considered
here, we have to assess the probability that $L+1$ independently selected vectors
across the hyper-surface of the sphere would happen to have prescribed values
of empirical correlation coefficients (within a small tolerance $\pm\epsilon$), because the
desired probability is equal to the ratio between the volume pertaining these
correlation coefficients and the total available volume. Using the techniques
of \cite[Proof of Theorem 5]{AM98a} and \cite{Merhav93}, it is not difficult
to show that the former volume is of the exponential order of
$\exp\{nh_G(X_0,X_1,\ldots,X_L)\}$, where $h_G(X_0,X_1,\ldots,X_L)$ is the
differential entropy of a Gaussian vector $(X_0,X_1,\ldots,X_L)$
with covariance matrix $\{\rho_{ij}S\}$ ($\rho_{ii}\dfn 1$, $i=0,1,\ldots,L$),
and the latter volume is of the exponential
order of $\exp\{\frac{n(L+1)}{2}\ln[2\pi eS]\}$.
Thus, the
probability under discussion is of the exponential order of
$\exp\{-nI_G(X_0;X_1;\ldots;X_L)\}$, where $I_G(X_0;X_1;\ldots;X_L)$ is the
multi--information associated with the above defined Gaussian vector $(X_0,X_1,\ldots,X_L)$,
in analogy to the finite--alphabet case. In particular,
let $\Lambda$ denote the $(L+1)\times(L+1)$ matrix of correlation coefficients
$\{\rho_{ij}\}$.
Then,
\begin{equation}
I_G(X_0;X_1;\ldots;X_L)=-\frac{1}{2}\ln|\Lambda|,
\end{equation}
and so,
\begin{equation}
E_{\mbox{\tiny ex}}^{\mbox{\tiny CKM}}(R,L)=
\inf_{\Lambda\in\calA(R)}
\left[\frac{LS}{2\sigma^2(L+1)^2}
\left(L+1-\frac{2}{L}\sum_{i<j}\rho_{ij}\right)-\frac{1}{2}\ln|\Lambda|-LR\right].
\end{equation}
where $\calA(R)$ is the set of all positive definite matrices
with $-\frac{1}{2}\ln|\Lambda|\le LR$.
For $L=1$, we get
\begin{eqnarray}
E_{\mbox{\tiny ex}}^{\mbox{\tiny CKM}}(R,1)=\inf_{|\rho|<1:~-\frac{1}{2}\ln(1-\rho^2)\le R}
\left[\frac{S}{4\sigma^2}
\cdot(1-\rho)-\frac{1}{2}\ln(1-\rho^2)-R\right]
\end{eqnarray}
which in the curvy part is attained with $\rho=\sqrt{1-e^{-2R}}$ to yield
\begin{equation}
E_{\mbox{\tiny ex}}^{\mbox{\tiny CKM}}(R,1)=\frac{S(1-\sqrt{1-e^{-2R}})}{4\sigma^2}
\end{equation}
and the affine part is the tangential straight line with slope $-1$.
For a general $L$, we argue that $E_{\mbox{\tiny ex}}^{\mbox{\tiny CKM}}(R,L)$
is always attained by a totally 
symmetric correlation matrix, i.e., $\rho_{ij}=\rho$ for all $i\ne j$, 
for some $\rho$ which keeps this matrix positive definite. This facilitates
considerably the minimization problem associated with $E_{\mbox{\tiny
ex}}^{\mbox{\tiny CKM}}(R,L)$, since it reduces to a minimization
over the single variable $\rho$. This total symmetry follows from the
concavity of the function $\ln|\Lambda|$ \cite[p.\ 679, Theorem 17.9.1]{CT06}.
In particular, if $\Lambda_0$ attains $E_{\mbox{\tiny
ex}}^{\mbox{\tiny CKM}}(R,L)$, then so does
$\Pi\Lambda_0\Pi^T$ for every permutation matrix $\Pi$ since neither the
objective function nor the constraint is sensitive to permutations. Now, let
\begin{equation}
\Lambda^*=\frac{1}{(L+1)!}\sum_{\Pi}\Pi\Lambda_0\Pi^T,
\end{equation}
which is obviously has a totally symmetric structure where all diagonal
elements are $1$ and all off--diagonal elements are the same. By the concavity
property,
\begin{equation}
\ln|\Lambda^*|\ge
\frac{1}{(L+1)!}\sum_{\Pi}\ln|\Pi\Lambda_0\Pi^T|=
\frac{1}{(L+1)!}\sum_{\Pi}\ln|\Lambda_0|=\ln|\Lambda_0|.
\end{equation}
It follows then that if $\Lambda_0$ is replaced by $\Lambda^*$, the value of
the objective function is reduced
without violating the constraint, and so, $\Lambda^*$ cannot be worse than $\Lambda_0$.
Now, for an $(L+1)\times(L+1)$ matrix $\Lambda$ whose entries are
$\rho_{ij}=\rho+(1-\rho)\delta_{i-j}$ ($\delta_k$ being the Kronecker delta
function), the eigenvalues are easily found to be $\lambda_1=1+\rho L$
(with an eigenvector being the all--one vector) and
$\lambda_2=\ldots=\lambda_{L+1}=1-\rho$ (with $L$ independent eigenvectors, all orthogonal to
the all--one vector). Therefore,
\begin{equation}
\ln|\Lambda|=\ln(1+\rho L)+L\ln(1-\rho).
\end{equation}
and so, the expurgated exponent function becomes
\begin{eqnarray}
E_{\mbox{\tiny ex}}^{\mbox{\tiny CKM}}(R,L)&=&\inf_{\{-1/L<\rho<1:~-\frac{1}{2}\ln(1+\rho
L)-\frac{L}{2}\ln(1-\rho)\le LR\}}
\left[\frac{SL(1-\rho)}{2\sigma^2(L+1)}-\right.\nonumber\\
& &\left.\frac{1}{2}\ln(1+\rho
L)-\frac{L}{2}\ln(1-\rho)-LR\right].
\end{eqnarray}
The behavior of this function can be characterized as follows.
Let $\rho_0$ be the unconstrained minimizer of the function
$$\frac{SL(1-\rho)}{2\sigma^2(L+1)}-\frac{1}{2}\ln(1+\rho
L)-\frac{L}{2}\ln(1-\rho)$$
over $[0,1]$ and let $\varrho(R)$ be the solution to the equation
$$-\frac{1}{2}\ln(1+\rho
L)-\frac{L}{2}\ln(1-\rho)= LR.$$
Then,
\begin{equation}
E_{\mbox{\tiny ex}}^{\mbox{\tiny CKM}}(R,L)=\left\{\begin{array}{ll}
\frac{SL[1-\varrho(R)]}{2\sigma^2(L+1)} & R<\varrho^{-1}(\rho_0)\\
\frac{SL(1-\rho_0)}{2\sigma^2(L+1)}-\frac{1}{2}\ln(1+\rho_0
L)-\frac{L}{2}\ln(1-\rho_0)-LR
& R\ge \varrho^{-1}(\rho_0)\end{array}\right.
\end{equation}
The critical rate, $R_0\dfn\varrho^{-1}(\rho_0)$, is the point at which the
derivative of the
function $SL[1-\varrho(R)]/[2\sigma^2(L+1)]$ w.r.t.\ $R$ is equal to $-L$, so that the
affine function in the second line of the last equation represents a straight
line that is tangential to
the curve of the first line at $R=R_0$. Such a point always exists because it
can easily be shown that the derivative of the curvy function begins from
$-\infty$ at $R=0$ and then increases. This is related to the fact that the
multi--information, as a function of $\rho$, has a zero derivative at $\rho=0$.

For $L=1$, we have already seen that $\varrho(R)=\sqrt{1-e^{-2R}}$. For $L=2$,
the derivation of $\varrho(R)$ is associated with the relevant solution of a cubic
equation in $\rho$, which is given by
\begin{equation}
\varrho(R)=\frac{\sqrt{1-e^{-4R}}}{2\cos\left[\frac{1}{3}\cos^{-1}(-\sqrt{1-e^{-4R}})\right]},
\end{equation}
with $\cos^{-1}(t)$ being defined as the unique solution $x$ to the
equation $\cos x= t$ in the range $[0,\pi]$.

\section{Future Work}

In Subsection 2.3, we have proved the well--known fact the optimal
list decoder provides the $L$ messages with highest likelihoods.
This proof suggests an
extension to a recent work \cite{Merhav13a} (see also
\cite{Merhav13b}) concerning
a decoder/detector that first has to decide whether the received channel
output $\by$ really contains a transmitted message or it is simply
pure noise (that is received when the transmitter is silent), and in the
former case to decode the message in the ordinary way ($L=1$). Three figures
of merit should therefore be traded off here: the false--alarm probability
(deciding that a message has been sent while $\by$ is actually pure noise),
the mis--detection probability (deciding that $\by$ is pure noise while it
actually contains a codeword) and the probability of decoding error (detecting
rightfully that $\by$ contains a codeword by decoding it erroneously).
It was shown in
\cite{Merhav13a} that the optimum decision rule in the sense of minimizing the
probability of decoding error subject to given constraints on the false alarm
and the mis--detection probabilities, is as follows: Accept $\by$ as containing
a message if and only if
\begin{equation}
e^{n\alpha}\sum_{m=1}^MP(\by|\bx_m)+\max_mP(\by|\bx_m)\ge
e^{n\beta}P(\by|\mbox{no~transmission}),
\end{equation}
where $\alpha$ and $\beta$ are constants that are chosen to meet the
false--alarm and the mis-detection constraints and
$P(\by|\mbox{no~transmission})$ is the probability of $\by$ when no
transmission takes place (e.g., the channel input is the zero signal), in other
words, the probability distribution of pure noise. The reason for the term
$\max_mP(\by|\bx_m)$ in this test turns out\footnote{See \cite{Merhav13a} for
details.} to be associated with the fact
that the probability of correct decoding
is proportional to
$\sum_{\by}\max_mP(\by|\bx_m)$.
Now, when list decoding is
considered, then in view of the last line of (\ref{Pc}), this decision rule is
generalized to the form:
\begin{equation}
e^{n\alpha}\sum_{m=1}^MP(\by|\bx_m)+\sum_{i=1}^LP(\by|\bx_{m_i^*(\by)})\ge
e^{n\beta}P(\by|\mbox{no~transmission}).
\end{equation}
The analysis of the error exponents associated with this detector/decoder can
be carried out using the same methods as in \cite{Merhav13a} for the fixed
list--size regime, but it is considerably more challenging in the exponential list--size
regime. 

\section*{Appendix}
\renewcommand{\theequation}{A.\arabic{equation}}
    \setcounter{equation}{0}

\noindent
{\it Proof of Theorem 2.}
For a given code $\calC$ and a given message $m$,
let $N_m(\hP,\calC)$ be the number of
$L$--tuples of distinct integers,
$\{m_1,\ldots,m_L\}$, all different from $m$,
for which $(\bx_m,\bx_{m_1},\ldots,\bx_{m_L})$ have a given joint empirical
distribution $\hP$ of an $(L+1)$--tuple of random variables all taking on values 
in $\calX$, whose single--letter marginals
(which are the individual empirical distributions of the various codewords)
all coincide with $Q$. 
Now, for the given code,
\begin{eqnarray}
P_{e|m}(\calC)&\le&\sum_{m_1,\ldots,m_L\ne m}\sum_{\by}
\left[P(\by|\bX_m)\prod_{i=1}^LP(\by|\bx_{m_i})\right]^{1/(L+1)}\nonumber\\
&=&\sum_{\hP}N_m(\hP,
\calC)\exp\{-n\hat{\bE}d(X_0,X_1,\ldots,X_L)\}.
\end{eqnarray}
Our key argument here is that
for every $\epsilon > 0$ and sufficiently large $n$,
there exists a code $\calC$ of rate (essentially) $R$,
that satisfies, for every message $m$ and every $\hP$,
\begin{eqnarray}
N_m(\hP,\calC)&\le&N^*(\hP)\nonumber\\
&\dfn&\left\{\begin{array}{ll}
\exp\{n[LR-\hI(X_0;X_1;\ldots;X_L)+\epsilon]\} & LR\ge
\hI(X_0;X_1;\ldots;X_L)-\epsilon\\
0 & LR< \hI(X_0;X_1;\ldots;X_L)-\epsilon\end{array}\right.
\end{eqnarray}
where $\hI(X_0;X_1;\ldots;X_L)$ is the empirical multi--information pertaining
to $\hP$.
To see why this is true, consider a random selection of the code $\calC$.
Then, obviously,
\begin{eqnarray}
\overline{N(\hP)}&\dfn&
\frac{1}{M}\sum_{m=0}^{M-1}\bE\{N_m(\hP,\calC)\}\\
&=&\bE\{N_0(\hP,\calC)\}\\
&\le&
M^L\cdot\mbox{Pr}\{(\bX_0,\bX_1,\ldots,\bX_L)\in\calT(\hP)\}\\
&=& M^L\cdot\frac{|\calT(\hP)|}{|\calT(Q)|^{L+1}}\\
&\exe& M^L\cdot\frac{\exp\{n\hH(X_0,X_1,\ldots,X_L)\}}{e^{n(L+1)H_Q}}\\
&=&\exp\{n[LR-\hI(X_0;X_1;\ldots;X_L)]\}.
\end{eqnarray}
It follows then that in the ensemble of the randomly selected codes
\begin{eqnarray}
& &\mbox{Pr}\bigcup_{\hP}\left\{\calC:~\frac{1}{M}\sum_{m=0}^{M-1}N_m(\hP,
\calC)> \exp\{n[LR-\hI(X_0;X_1;\ldots;X_L)+\epsilon/2]\}\right\}\nonumber\\
&\le&\sum_{\hP}\mbox{Pr}\left\{\calC:~\frac{1}{M}\sum_{m=0}^{M-1}N_m(\hP,
,\calC)> \exp\{n[LR-\hI(X_0;X_1;\ldots;X_L)+\epsilon/2]\}\right\}\nonumber\\
&\le&\sum_{\hP}\frac{\overline{N(\hP)}}
{\exp\{n[LR-\hI(X_0;X_1;\ldots;X_L)+\epsilon/2]\}}\nonumber\\
&\lexe&\sum_{\hP}e^{-n\epsilon/2}\nonumber\\
&\le&(n+1)^{|\calX|^{L+1}}\cdot e^{-n\epsilon/2}\to 0,
\end{eqnarray}
which means that there exists a code (and in fact, for almost every code),
\begin{equation}
\frac{1}{M}\sum_{m=0}^{M-1}N_m(\hP,
\calC)\le \exp\{n[LR-\hI(X_0;X_1;\ldots;X_L)+\epsilon/2]\}~~~\forall
\hP.
\end{equation}
For a given such code and every given $\hP$, there must then
exist
at least $(1-e^{-n\epsilon/2})\cdot M$ values of $m$ such that
\begin{equation}
N_m(\hP,
\calC)\le \exp\{n[LR-\hI(X_0;X_1;\ldots;X_L)+\epsilon]\}.
\end{equation}
Upon eliminating the exceptional codewords from the code, for all
$\hP$, one ends up with at least
$[1-(n+1)^{|\calX|^{L+1}}e^{-n\epsilon/2}]\cdot M$ for which
\begin{equation}
N_m(\hP,
\calC)\le \exp\{n[LR-\hI(X_0;X_1;\ldots;X_L)+\epsilon]\}~~~~\forall
\hP.
\end{equation}
Let $\calC'$ denote the sub--code formed by these
$[1-(n+1)^{|\calX|^{L+1}}e^{-n\epsilon/2}]\cdot M$
remaining codewords. Since $N_m(\hP,
\calC')\le N_m(\hP,
\calC)$, then the sub--code $\calC'$ certainly satisfies
\begin{equation}
N_m(\hP,
\calC')\le \exp\{n[LR-\hI(X_0;X_1;\ldots;X_L)+\epsilon]\}~~~~\forall
\hP.
\end{equation}
Finally, observe that since $N_m(\hP,
\calC')$ is a non--negative integer, then for $\hP$ with
$LR-\hI(X_0;X_1;\ldots;X_L)+\epsilon< 0$, the last inequality means
$N_m(\hP,
\calC')=0$, in which case the r.h.s.\ of the last equation becomes
$N^*(\hP)$. Thus, we have shown that there exists a code
$\calC'$ of
size $M'=[1-(n+1)^{|\calX|^{L+1}}e^{-n\epsilon/2}]\cdot e^{nR}$ for which all
codewords satisfy $N_m(\hP,
\calC')\le N^*(\hP)$ for all $\hP$.

As a consequence of the above observation, we have seen the existence
of a code $\calC'$ for which
\begin{eqnarray}
& &\max_m P_{e|m}(\calC')\nonumber\\
&\le&\sum_{\hP}N^*(\hP)
\exp\{-n\hat{\bE}d(X_0,X_1,\ldots,X_L)\}\\
&=&\sum_{\hP: N^*(\hP)> 0}
\exp\{n[LR-\hI(X_0;X_1;\ldots;X_L)-\hat{\bE}d(X_0,X_1,\ldots,X_L)+\epsilon]\}
\end{eqnarray}
and due to the arbitrariness of $\epsilon> 0$, this means (upon maximization
over $Q$) that there exists a
sequence of rate--$R$ codes for which
\begin{equation}
\lim_{n\to\infty}\left[-\frac{\ln \max_m P_{e|m}(\calC)}{n}\right]\ge
E_{\mbox{\tiny ex}}^{\mbox{\tiny CKM}}(R,L),
\end{equation}
where $E_{\mbox{\tiny ex}}^{\mbox{\tiny CKM}}(R,L)$ is defined as in Theorem 2.
This completes the proof of Theorem 2.

\end{document}